\documentclass[galaxies,review,accept,pdftex,oneauthor]{Definitions/mdpi} 

\firstpage{1} 
\pubvolume{1}
\issuenum{1}
\articlenumber{0}
\pubyear{2025}
\copyrightyear{2025}
\externaleditor{Firstname Lastname}
\datereceived{4 May 2025 } 
\daterevised{31 May 2025 } 
\dateaccepted{12 June 2025 } 
\datepublished{ } 
\hreflink{https://doi.org/} 

\Title{Red Supergiant Mass Loss and Mass-Loss Rates}

\TitleCitation{Red Supergiant Mass Loss and Mass-Loss Rates}

\Author{Jacco Th. van Loon \orcidA{}}
\AuthorNames{Jacco Th. van Loon}

\AuthorCitation{van Loon, J.Th.}
\address[1]{Lennard-Jones Laboratories, Keele University, ST5 5BG, UK; j.t.van.loon@keele.ac.uk}
\abstract{This review discusses the causes, nature, importance and observational evidence of mass loss by red supergiants. It arrives at the perception that mass loss finds its origin in the gravity which makes the star a star in the first place, and is a mechanism for the star to equilibrate. This is corroborated by a careful examination of various popular historical and recent empirical mass-loss rate prescriptions and theoretical works, and which provides no evidence for an explicit dependence of red supergiant mass loss on metallicity though dust-associated mass loss becomes less prevalent at lower metallicity. It also identifies a common problem in methods that use tracers of mass loss, which do not correct for varying scaling factors (often because there is no information available on which to base such correction) and as a result tend to underestimate mass-loss rates at the lower end. Conversely, dense, extended chromospheres in themselves do not translate into high mass-loss rates, and the significance of stochastic mass loss can be overstated. On a population scale, on the other hand, binary interaction acts as a stochastic agent of mass loss of great import. In all, evidence is overwhelming that points at red supergiants at the lower mass end losing mass at insufficient rates to shed their mantles before core collapse, but massive (at birth) red supergiants to be prone to intense, dusty mass loss which sees them become hotter stars before meeting their fate. This is consistent with the identified progenitors of hydrogen-rich supernovae. Supernova evolution holds great promise to probe the mass loss but we caution against confusing atmospheres with winds. Finally, promising avenues are looked into, which could forge step-change progress in what has been a long and arduous search for the holy grail of red supergiant mass loss.  We may yet find it!}


\begin{document}

\section{Prologue}

This review is not meant to compare all, and~recommend a ``best'' mass-loss recipe. Data limitations and choices of input and parametrisation still present obstacles to such judgment. Instead, it aims at identifying general trends, both in terms of observed behaviour and observational and analytical biases, and~examines the way in which mass loss is being expressed, in~order to focus our minds when moving ahead. There will therefore be important aspects of the atmospheres and winds of red supergiants that are being studied for very good reasons, but~that are not addressed here in full. We refer the reader to the excellent 2021 review by Leen Decin 
\cite{Decin2021} for much of what complicates deriving and predicting mass-loss rates, and~here make an attempt to reduce the problem to its simple roots and its pertinent outcomes. It starts with a general reflection on what stellar mass loss is, why it happens, and~how this looks more specifically for red supergiants, before~we shall delve into the (wealth of) observational evidence and (limited) theoretical effort where most of the references to the literature can be found. Throughout, we have tried to recover where ideas were coined for the first time, in~an effort to preserve an accurate and inspiring historic credit record in an information age where this is easily lost. We end with a synthetic extraction of salient points, and~a selection of recently available powerful methods of which some caveats and pitfalls are highlighted. We must embrace these capabilities but never forget to also apply our own mind to~it.

\section{An Introduction to Stellar Mass~Loss}

It is a fact that all stars lose mass, no matter their size, age or composition. Stars lose mass as soon as they are born and they continue to lose mass throughout their lives. Indeed, mass loss is not reserved for the dying but it is seldomly acknowledged as a necessity of being a star. All stars, without~exception, lose mass as a result of intrinsic processes by means of two distinct agents: radiation and particles. While we shall focus on the latter, it pays to reflect on the former, too.

Eventually, photons emitted by a star carry away most of the mass defect arising from nuclear fusion reactions that occur deep inside the star (neutrinos carry away the remainder). While the resulting luminosities may seem impressive, no star loses more than about a thousandth of its initial mass in this way. We shall thus not concern ourselves with this source of mass loss. Yet, particulate mass loss cannot be seen in isolation from radiative losses, causally as well as both being based on common fundamental~principles.

Firstly, the~photons that leave the star are only indirectly produced by nuclear fusion. In the 
first instance, the~kinetic energy of the fusion products along with gamma rays, heats up the stellar plasma to create the pressure gradient that counteracts gravity and keeps the star stable. The~layer that heralds the transition between the stellar interior where radiation and matter are coupled, and~the exterior space where they decouple, has a physical non-zero temperature and thus radiates. Photons emitted outward from this photosphere are lost as they exceed the local escape speed. They can still interact with particles further out, though, heating them and imparting outward-directed momentum onto them. If~this medium is also optically thick then a second photosphere envelopes the star and this is what we as a distant observer would be looking~at.

The huge temperature gradient that accompanies the pressure gradient and facilitates the transport of energy outwards also results in a stratified ionisation structure, with~associated variations in opacity. In~some cases this can lead to oscillatory radial motion of the stellar mantle due to the alternation between excess radiation pressure and surplus weight---the so-called ``kappa'' mechanism 
\cite{Baker1965}. However, energy transport is often not fast enough without bulk plasma motion carrying atomic kinetic energy upwards to near the photosphere where it is deposited, after~which the cooled gas falls back. This convection alters the structure of the star markedly, both physically and in terms of mixing. It can also act mechanically, exciting internal oscillations that can manifest in the form of radial pulsation, driving external acoustic waves, 
or generate and interact with electromagnetic 
fields. All of this is a consequence of the sustained conversion of mass into other forms of energy through nuclear~fusion.

Secondly, nuclear fusion inside stars only happens because gravity created the extreme conditions for it. The~uni-directional curvature of space-time caused by mass naturally leads to collapse, which increases density but also atomic kinetic energy. Whether it is in the core or in a shell around it, nuclear fusion is a self-regulatory response of a star to its own gravity. For~as long as it lasts. Typically, core hydrogen burning is most efficient and least demanding, making for the least luminous and therefore longest-lasting 
phase in the life of a star. Switching to helium burning increases the temperature requirements, which drastically increases the burning rate and hence the luminosity, curtailing the lifetime. It also reconfigures the star, exacerbating the contrast between the dense core and extended mantle. This trend continues for any further switches in nuclear burning, but~these become so short in duration that it is unclear whether they would alter the outer boundary conditions. Likewise, transitions to shell burning are associated with similar increases in internal contrast compared to core burning. Red supergiants are generally core helium burning, but~it is possible that a small fraction are caught burning carbon 
\cite{Woosley2002}.

Particulate mass loss, hereafter simply referred to as the ``mass loss'', typically takes the form of a wind, at~least at some height above the stellar photosphere. Averaged over sufficient scales in time and space these winds are steady but they change as a star evolves. Table~\ref{tab1} summarises some of the nomenclature in order to provide clarity and context. Stellar winds tend to increase in speed and strength with luminosity, resulting in higher mass-loss rates and ram pressures even if radiation pressure is not their driving mechanism. This suggests that luminosity may act as a proxy for the true underlying physical reason for the mass loss. This suspicion is corroborated by the more complex behaviour of the wind speed, which tends to be slower for red supergiants than for hotter stars. In~fact, both mass-loss rate and wind speed may be more closely tracing the surface gravity, where ``surface'' can be rather ill defined for red supergiants. This is intriguing as it could point at a more intimate relation between mass loss and gravity, even if the wind may not yet have reached the escape speed and thus require more prolonged driving (not necessarily accelerating it, but~at least pushing it until it would escape).

\begin{table}[H] 
\caption{Overview of forms of boundary layer flow pertaining to cool~supergiants.\label{tab1}}

\begin{adjustwidth}{-\extralength}{0cm}

\begin{tabularx}{\fulllength}{lLL}
\toprule
\textbf{Form}			& \textbf{Common Meaning}
					& \textbf{Characteristic Scales}								\\
\midrule
Convection			& Buoyant upwelling and fallback cycle
					& $\sim$10$^{-1}$\,R$_\star$ with roll timescale $\sim$yr			\\
Pulsation				& Global radial oscillation of outer mantle
					& $<$R$_\star$ with periods of a few~yr						\\
Rotation				& Global azimuthal oscillation of mantle
					& $>$R$_\star$ with periods $>$10\,yr						\\
Luminosity			& Photonic energy loss rate
					& $10^{4-5.5}$\,L$_\odot$	 for $10^{5\pm1/2-3}$\,yr (He/C)		\\
Mass loss	(ML)			& Matter lost by the star permanently
					& few $10^{-1}$\,M$_\star$ at $10^{-5.5\pm2}$\,M$_\odot$\,yr$^{-1}$\\
Episodic ML			& Temporally abrupt varying mass loss
					& $10^{3\pm1}$\,yr										\\
Eruptive ML			& Sudden sporadic ejection of matter
					& $\Delta$t$\sim$$10^{0-1}$\,yr, $\Delta M\gg\dot{M}_{\rm before}\Delta t$\\
Stochastic ML			& Irregular ejection of matter \textsuperscript{1}
					& $\sim$yr\\
Wind					& Continuous flow by which mass is lost
					& wind speed$\sim$20--100\,km\,s$^{-1}$					\\
Acoustic waves			& Longitudinal pressure oscillation
					& sound speed of few km\,s$^{-1}$							\\
Alfv\'en waves			& Transverse ion--magnetic field oscillation
					& $<$m (heating) -- $0.1$\,R$_\star$ (driving)					\\
Shocks				& Supersonic displacement of matter
					& $>$10\,km\,s$^{-1}$									\\
\bottomrule
\end{tabularx}
\end{adjustwidth}
\noindent{\footnotesize{\textsuperscript{1} Usually both in a temporal and spatial sense, not necessarily exceeding the temporally \& spatially averaged ML.}}
\end{table}

The actual mechanism that lifts matter off the star can be diverse and multi-faceted. It takes place around the stellar boundary layer, but~can originate underneath it and continue and transform outside it. The~Sun's wind flows out of its hot corona (Parker 
~\cite{Parker1960,Parker1965}), but~only because the corona is heated by magnetic field reconfigurations and possibly Alfv\'en waves that find their origin in plasma motions in the chromosphere, itself a superheated layer above the photosphere where convection drives acoustic waves. Often more eruptive mass ejections occur as well, at~or near the base but propagating throughout the wind. Likewise, in~luminous cool red giant stars, 
convection cells and global radial pulsation increase the scale height of the already extended atmosphere permanently but in a time and spatially varying manner. All these processes, complex and dynamic though they may be, ultimately are instantiated by the simple workings of gravity. Mass loss can be regarded as a natural consequence of a quasi-static state whereby the star continuously approaches but never quite is in a state of equilibrium. A~star is not an isolated system, and~the boundary phenomenon that is mass loss does not have the same significance in the grand scheme of the~Universe.

Some stars also lose particles caused or exacerbated by extrinsic triggers, which usually involves a companion star, planet or stellar remnant (or in rare cases, the~tidal disruption upon a close encounter with a supermassive black hole in the centre of a galaxy). This may be due to a variety of reasons, from~the heating and pressure effects of irradiation, through ram pressure exerted by the companion's own wind (or explosive expulsion), to~the tidal imbalances between the gravitational field and angular momentum distribution. What all of these scenarios have in common is that they break the spherical or axial symmetry inherent in the case of a single gravitating object. Just like rotation of a star breaks spherical symmetry and alters plasma flow directions (with implications for internal mixing, oscillation, convection and magnetic field strength and topology), so do the gravitational perturbations arising from the presence of another body of mass modify the boundary through which mass is lost. These complications move the system further away from equilibrium and thereby enhance the mass loss, but~they do not in themselves cause mass loss except if the system becomes partially unbound when rotating at break-up speed, overflowing through the inner Lagrangian point or merging into a common~envelope.

In summary, all stars lose mass because they are not in equilibrium with the rest of the Universe. There are two equilibrium configurations: a singularity or a uniform distribution. A~star occupies an intermediate and therefore only temporary state. While gravity moves it towards the first solution, in~accordance with the second law of thermodynamics and the growth in entropy when more probable states are attained, work is done and heat is expelled and this takes the form of radiation and mass loss. Only in the extreme case of an event horizon is this avoided. Gravity sets the radius of the nuclear burning zone and, through entropy maximisation driven by the generated luminosity, the~outer, cooler boundary location (``stellar radius''), the~latter being limited by the point at which the stellar atmosphere becomes transparent and radiation and matter decouple. Hence ``surface'' gravity is expected to play a decisive r\^ole in setting the mass-loss rate in comparison to the luminosity. These are energetically comparable for the Sun ($\dot{M}c^2/L$$\sim$1), but in red supergiants mass loss far exceeds radiation~loss.

\section{Why and How Red Supergiants Lose Mass and Why It~Matters}

We shall now focus on the specific case of red supergiants. These were once luminous mass-losing core hydrogen burning stars, before~core helium burning ignited and their mantles expanded in response, cooling their photospheres (not their cores!) on the rapid thermal timescale that has so far eluded direct observation. Because~the preceding winds were much faster, the~red supergiant wind expands into a vacuum. However, their modest age of order 10 million years means they are more likely than an average star to be found in the vicinity of young(er) massive stars with their ionising radiation fields and fast stellar winds that can impact upon the red supergiant's own, especially if the red supergiant is a member of a star cluster 
\cite{Larkin2025}. What is less expected is interaction between their wind and dense molecular clouds, since these will have been dispersed at least locally by prior stellar feedback~already.

While both red supergiants and their progenitors, and~the Sun for that matter, lose mass for the same fundamental reasons as we have just discussed, their winds are very different in nature. Red supergiant photospheres are much cooler, around 4000\,K, than~those of their blue ancestry, above~10,000\,K. The~latter, O- and B-type stars possess the fortunate combination of sufficient ionisation and sufficient ultraviolet light to efficiently couple the radiation field to the upper atmosphere through line opacity scattering 
\cite{Castor1975}. This is not the case for red supergiants. The~warmer cases among them, of~spectral K-type, have mildly ionised or neutral atomic atmospheres that provide very little opacity for radiation pressure to significantly reduce the effective gravity. Their cooler counterparts, of~spectral M-type, allow molecules to form in abundance, especially H$_2$, CO, water and other oxides such as TiO, and~these provide opacity across the optical and near-infrared where the photosphere is brightest. However, radiation pressure is limited because on the one hand the molecular absorption bands comprise a blend of transitions that cast a shadow on any light that could have become available as the transition Doppler shifts upon acceleration of the molecule (such as is the case in the strong resonance lines of hot star winds), while on the other hand the molecular bands leave sufficient transparency in between them for light to escape~unscathed.

The coolest red supergiants are most prone to strong radial pulsations, near-sonic but operating through cycles of years, that create long-lived dense layers high up in an otherwise highly diluted (hence not in local thermodynamic equilibrium) atmosphere, shocked by the impact of back-falling matter. This promotes the efficient condensation of gas onto molecular nucleation seeds such as titanium and aluminium oxides and their solidification into (sub)micrometre-sized 
grains composed mainly of minerals of the silicate family with iron and magnesium inclusions. This circumstellar dust provides continuum opacity that is relatively high in the optical where the radiation field carries plenty of momentum. The~accelerating grains undergo elastic and inelastic collisions mostly with atomic and molecular hydrogen, whereby net outward momentum is imparted onto the gas. The~coupling between the dust and gas fluids is not in general stiff, causing the gas to lag behind and the grains to flow through the gas with a relative drift speed. It is important to note that dust formation never completely depletes the gas of its refractory elements, and~that in warmer red supergiants (early-M type) the dust fraction and grain sizes are much smaller, radiation pressure on dust is generally unable to drive the wind and dust is instead dragged along by the gas. What drives the winds of warm red supergiants (and yellow hypergiants) is yet to be established, despite the fact that most red supergiants are too warm for strong radial pulsation and dusty~winds.

It should be emphasised how extensive and rarefied the boundary layers of red supergiants really are. With~sizes in excess of Jupiter's orbit (though not all are that big) their mantles have densities many orders of magnitude below that of air. They are only opaque because of their sheer size and thus path length crossing it, but~the transition from optically thick to optically thin takes place over many solar radii and indeed, red supergiants have markedly different sizes when seen in molecular absorption bands (larger) or in between them (smaller). With~the speed of sound, convection and pulsation all in the region of several kilometres per second, the~dynamical timescales are months for localised convection cells and years for global pulsation. The~escape speed still exceeds these by a factor of a few, hence ballistic launches alone do not result in mass loss. Winds of red supergiants are typically seen to be accelerated until they reach escape speed at substantial distance from the photosphere where the escape speed has dropped to around 10\,km\,s$^{-1}$. This corresponds to that at Earth's surface, but~the gravitational acceleration even at the red supergiant's photosphere is several orders of magnitude lower than that at Earth's surface—this implies that a strong force is not necessarily what is required, leaving many options on the table, but~that they would need to operate over sufficient duration and thus distance in order to reach the required speed. Radiation pressure does this quite well, diminishing according to the inverse distance square law just as gravity does, unless~optical depth is significant. Ram pressure and acoustic waves (acting longitudinally) or electromagnetic Alfv\'en waves (acting transversally), on~the other hand, readily dissipate energy 
\cite{Hartmann1980}. While this is a requirement for accelerating matter these waves thereby also damp out quickly—the reason pulsation shocks emit in H$\alpha$ and how the Sun's corona is heated to a million~degrees.

And thus, nature adjusts. When pressure gradients become too shallow to drive a wind, radiation pressure or wave deposition takes over. When that fails, akin to the switch in mode of energy transport through the stellar interior, bulk motion in the form of convection and especially pulsation grow in importance. The~situation is more complex and notably confusing for red supergiants because of the different regimes encountered both among individual stars and across their own extended atmospheres. Despite ranging by less than a factor of two, the~most sensitive discriminator is not mass or luminosity but stellar photospheric temperature, masquerading metallicity as a compounding factor since photospheric temperatures tend to be (weakly) anti-correlated with metallicity because of its effect on the opacity and therefore extent of the stellar mantle. This makes red supergiants unique in being such specific stellar class and yet internally hugely disparate, with~completely different mass-loss and wind-driving mechanisms at the warmer and cooler ends of the spectrum. The~coolest red supergiants are good examples where the mass-loss mechanism likely requires more than a single process, one operating near the base (pulsation) and another at arm's length (radiation pressure on dust). In~the end, this must be explained in terms of the variations in depth and steepness of the gravitational potential—the curvature of~space.

Within the context of understanding the physical conditions at the boundary layer and their relevance to mass loss, we must consider rotation, or~azimuthal motion in general since angular momentum affects effective gravity and flow direction. Often dismissed because even main-sequence stars rotating near break-up speed are expected to slow down to red-supergiant surface speeds of no more than a few km\,s$^{-1}$, this assumes no angular momentum is transported from the inside out (for instance by magnetic tension), no torque is applied by a close companion, and~that even a mere kilometre per second is negligible. But~this is within the same order of magnitude as the speed of sound, pulsation and convection, and~thus can lead to Coriolis deviations away from pure radial motion, modified oscillation resonance patterns and fluid instabilities and turbulence with possible implications for magnetic field reconnections, wave propagation, molecular chemistry and so on. Convection motions across the stellar surface (from their warm uprising centres towards their down-drafting edges) may exceed even the most optimistic rotational speeds and induce similar effects. This naturally complicates the picture, but~need not prevent us from accounting for it in the assessment of red supergiant mass~loss.

Finally, there is strong evidence for binary interaction to be prevalent among massive stars. Binary (and in fact triplet) fractions among main-sequence and slightly evolved massive stars are high, but~importantly, their separations are often relatively small and their mass ratios often close to unity 
\cite{Sana2012,Frost2025}. This probably reflects angular momentum considerations during their formation but also means that as these stars evolve and expand they are likely to transfer mass or come into contact, of~which VV\,Cephei is a good example, and~possibly merge as is suspected to have been the case in luminous red novae such as V838\,Monocerotis 
\cite{Kaminski2021}. If~the more massive star in the system meets its fate in the form of a supernova in which copious mass is ejected then the system can become unbound, giving the other star a high space motion, or~result in a compact binary of which one component is a neutron star or a black hole. The~latter can become a high-mass X-ray binary when the compact remnant accretes mass from the stellar companion. If~that happens through Roche lobe overflow then this will reduce the mass of the stellar mantle. Obviously this will affect its subsequent evolution and likely prevent it from becoming a red supergiant. But~if the accretion is wind-fed, the star may become a red supergiant and possibly engulf the compact remnant, becoming a Thorne—\.{Z}ytkow object 
\cite{Thorne1975}. It remains unclear in what way such product can be distinguished from an ordinary red supergiant, and~how its mass loss would be~affected.

Hydrogen-poor core-collapse supernovae are generally understood to have followed the stripping of the stellar mantle as a result of binary interaction; however, there has to be a fraction of progenitors of which the mantle was depleted through a stellar wind or intrinsic eruptive event. All confirmed progenitors of hydrogen-rich supernovae have been red supergiants (type II-P) or yellow hypergiants (type IIb), but~none of them appeared to have been more massive than about 20\,M$_\odot$ at birth 
\cite{Smartt2009,Smartt2015}, which leaves the more massive red supergiant descendants unaccounted for. It is unlikely that all of them were stripped or prevented from becoming red supergiants in the first place due to binary interaction because some will have been single stars or part of a sufficiently wide system to have essentially evolved as a single star. Either the (initially) most massive red supergiants seriously deplete or blow off their mantles by their own mass loss before core collapse ensues, which could help explain the populations of hydrogen-poor massive Wolf--Rayet stars and yellow hypergiants such as IRC\,10$^\circ$420, or~they envelop themselves in so much dust that they have not been recorded in pre-supernova images and thereby eluded discovery, not helped by the fact that these specimens are a lot rarer than their less massive counterparts and therefore harder to be caught in the act of exploding (or they implode without \mbox{supernova 
\cite{Kochanek2008}}). Either way, this makes strong statements about red supergiant mass loss, and~possibly requires 
a strong positive dependence on birth mass or current space curvature. This could be a self-fulfilling prophecy, as~mass loss leads to mantle expansion (essentially the star becomes less bound), maybe enhancing further mass loss but certainly facilitating dust production. The~one and only clear-cut case we have of a supernova arisen from the death of a blue supergiant, SN\,1987A revealed a ring of dust likely produced during a previous incarnation as a red supergiant some $10^4$ years before the core collapsed (see~\cite{McCray1993} for an early review). 
Most scenarios that have been proposed to explain this involve a stellar merger, a~concept which is in fashion, but~whatever really happened it is clear that the mass loss from the red supergiant was non-isotropic with strong axial symmetry, and~had led to the removal of almost the entire~mantle.

Red supergiants do not present nucleosynthetic surface enrichment to the same extent that lower-mass stars do as they ascent the asymptotic giant branch (analogous to the shorter red supergiant branches and longer first ascent red giant branches), aside from an expected nitrogen enhancement. The~exact time dependence of the mass-loss rate is therefore less important for this reason. However, while mass-loss rates of lower-mass stars undercut nuclear (and hence evolutionary) timescales and thus truncate evolution by exposing the stellar core, the~situation is much more complex for red supergiants. Mass loss does terminate the red supergiant phase before core collapse takes place for some stars (thereby metamorphosing into yellow hypergiants and Wolf--Rayet stars), but~not for many others (which are known to explode as hydrogen-rich supernovae). Interestingly, the~super-AGB stars, the~most massive asymptotic giant branch stars, that ignite core carbon burning but present shell burning flashes (thermal pulses), may exhibit a similar mixed bag of outcomes, some leaving massive oxygen—neon white dwarfs whilst others undergo electron-capture supernovae 
\cite{Doherty2017}. The~vast majority of red supergiants are losing mass at rates below that of the (high) nuclear mass consumption rate via the triple-$\alpha$ core helium burning process. Some are clearly seen to be losing mass at a much higher rate. The~question is unanswered as to how long this lasts and hence what consequences it has for the star's fate, and~which red supergiants enter this phase—as clearly not all red supergiants do. As~discussed, it appears that it may only be the red supergiants from the most massive progenitors that exhibit this superwind phase, perhaps owing to the lower surface gravity (higher luminosity by proxy) and/or more substantial reduction in mantle mass by preceding blue giant mass loss; mass-dependent outcomes of interacting binary evolution may play a r\^ole, but~if this were the only mechanism then some massive red supergiants would be expected that would never experience the superwind~phase.

While the end products of massive star evolution provide constraints on the red supergiant mass-loss history, more importantly the mass-loss history of red supergiants determines their end products. If~the star loses enough mass then its mantle will be depleted and its 
underlying layers exposed. If~not, then it will meet its fate as a hydrogen-rich supernova. The~amount of mantle left determines the rise and shape of the supernova afterglow light curve, and~the amount of explosive nucleosynthesis that takes place and thus the chemical enrichment of interstellar space. Mass lost in the preceding years, decades and centuries can be gleaned from the spectral changes in the supernova afterglow and affects the early evolution of supernova remnants. There is evidence for dense circumburst gas in many core-collapse supernovae which has been interpreted as either a brief episode of intense mass loss shortly before core collapse or the retention of material in the vicinity of the star due to binary gravitational effects. If~mass loss somehow "knows" of the impending death, this requires a mechanism to connect the unfolding events in the deep interior to the outer boundary layer. It would also mean that such eruptions could herald an imminent supernova and thus provide an early warning sign that could be searched for in anticipation of the~explosion.

Supernovae, owing to their effective conversion of hydrogen and helium to refractory metals are prolific dust factories. However, their blast wave and the forward shock that follows it can sputter pre-existing dust formed in the red supergiant wind, and~the circumburst density sets the conditions for the strength of the reverse shock which goes on to sputter a substantial fraction of the supernova dust 
\cite{Lakicevic2015}. Given the importance of supernovae as potential sources of cosmic dust at high redshift there is much interest in finding out whether the reverse shocks are stronger or weaker at those epochs. Strong but slow red supergiant winds, for~instance, could lead to lower supernova dust (effective) yields, but~it is also conceivable that the red supergiant phase is shorter or absent in the primitive Universe, in~which case the weaker, faster hot star winds allow for much larger fractions of supernova dust to enter the interstellar medium of pristine~galaxies.

\section{Measurements of Mass-Loss Rates and Lessons (To Be) Learnt}

While mass-loss rates in themselves are meaningless without knowing the timescale over which they persist and hence how much mass is lost in total, they are directly linked to the mechanisms by which the mass is lost and the conditions under which it unfolds. Besides~the rates of mass loss, other aspects of the mass loss can be pursued such as the dynamics, spatial and temporal variation and correlation (if not causal relation) with stellar parameters such as luminosity, temperature, wind composition or indeed metallicity and so on. Likewise, the~way in which mass loss is seen or indeed measured reflects the way in, and~means by which it happens, though~as we shall see this could also be misleading. It would be extremely helpful if effects of binarity could be isolated, as~this separates stars with otherwise similar expectations for their mass loss and fate. Unfortunately, statistical studies of the latter kind are in their infancy (except for stripped supernova progenitors) and our knowledge of how binarity changes mass-loss rates is largely anecdotal. We shall here focus on the way in which mass-loss rates have traditionally been determined, and~what we have been learning from this, in~the 
first instance thinking in terms of presumed canonical evolution of single red supergiants. We will look at case studies to examine the validity of these~premises.

That said, the~field of study of red supergiant winds started in earnest when Armin Deutsch in 1956 published an 
analysis of neutral and singly-ionised metallic absorption lines arising in the extended envelope of the nearby (110\,pc) M5\,Ib--II star Ras Algethi ($\alpha$\,Herculis) but seen also in the spectrum of its visual G8-type companion 
\cite{Deutsch1956}. Violet-asymmetry in the line profiles in the atmosphere indicated a launch speed around 10\,km\,s$^{-1}$, which puzzled Deutsch given the significantly higher escape speed from the stellar surface and it was proposed that some unknown mechanism provided sustained acceleration to counter gravity. Uncertainties in model assumptions precluded from determining a precise mass-loss rate but a minimum value of $3\times10^{-8}$\,M$_\odot$\,yr$^{-1}$ was presented. It was noted that similar violet-asymmetric lines had been seen in other cool luminous giants including Betelgeuse ($\alpha$\,Orionis) and Antares ($\alpha$\,Scorpii, which also has a visual companion but of B-type hence affecting the ionisation balance in the envelope). While intriguing, $\alpha$\,Herculis is not a red supergiant but an asymptotic giant branch star of about 3\,M$_\odot$ 
\cite{Moravveji2013} but more seriously, reliance on atmospheric line profiles for mass-loss rate measurements is complicated by the fact that atmospheric motions, outward and inward, also result from pulsation (and convection) alone and do not necessarily indicate mass loss. Indeed, Deutsch noted that in some cases, red-asymmetries had been seen and that early-M-type (super)giants exhibit chromospheres, and~H$\alpha$ emission, too, can arise in an otherwise static atmosphere 
\cite{Dupree1984} or signal pulsation shock dissipation. Some of the more appropriate, resonance lines to use occupy the ultraviolet, such as the Mg\,{\sc ii}\,2798 doublet, but~although detected in the atmosphere of Betelgeuse, late-type giants tend to be faint at these wavelengths and the near-infrared line of the abundant helium, He\,{\sc i}\,10,830 may be more suitable—see the excellent 1986 review by Andrea Dupree 
\cite{Dupree1986}.

Possibly the most cited publicly unavailable (hence rarely read) work, the~seminal 1975 presentation of an empirical formula to capture the mass-loss rates of cool luminous stars by Reimers 
\cite{Reimers1975} has been widely used in theoretical stellar evolution tracks: $\dot{M}=\eta L/gR$ (all solar units and per year). The~constant $\eta=4\times10^{-13}$ has since been considered to be adjustable, and~some have done so arbitrarily to suit their needs. The~form of Reimers' law remains interesting as it suggests a relation with the photon energy loss rate ($L$) and the square of the escape speed ($v_{\rm esc}^2=2gR$), interpreted by 
\cite{Dupree1986} in terms of a constant fraction of the luminosity being used to increase the specific potential energy of escaping matter. Reimers 
\cite{Reimers1977} then found that the wind speed measured from chromospheric Ca\,{\sc ii}\,H+K line shifts also scales with the square of the escape speed, falling short of escape for luminous cool giants but on a par for the case of the Sun. Assuming, then, that the material seen lifted off the photosphere does indeed escape, it still requires a mechanism to make that happen in addition to the chromospheric dynamics alone. Reimers suggested a ``cool corona'', perhaps not quite acting in the same way as for the thermally-driven solar wind but taking the form of Alfv\'en waves as developed subsequently by Hartmann \& MacGregor 
\cite{Hartmann1980} and others 
\cite{Charbonneau1995,FalcetaGoncalves2006}. The~problem is exacerbated as the canonical Reimers law underestimates the mass-loss rates measured for the coolest, most luminous red~supergiants.

More recently, Schr\"oder \& Cuntz 
\cite{Schroeder2005} proposed a physically motivated modification of the ``constant'' $\eta$ to account for the effect of the chromosphere on the atmospheric structure and dynamics. This introduces an explicit temperature dependence ($\propto T_{\rm eff}^{3.5}$) and an additional gravity ($g$) dependence which in practice kicks in at the very lowest gravities (towards $\propto 1/g$) and vanishes for main sequence stars. For~red supergiants it would tend towards $\dot{M}\propto LR^2T_{\rm eff}^{3.5}/M^2\approx (L/M)^2$. Their modified form of Reimers' law better reproduces the measured mass-loss rates for red supergiants than the original \mbox{form 
\cite{Schroeder2007}} though, incidentally, both overestimate the mass-loss rate of the Sun by a factor of about~ten.

Meanwhile, back in 1971, Gehrz \& Woolf 
\cite{Gehrz1971} had found that some red supergiants display excess infrared emission due to circumstellar dust, as~do many asymptotic giant branch stars. By~assuming that dust will have condensed at distances where the equilibrium temperature has dropped below the sublimation point of silicates they could determine circumstellar masses, and~by adopting the typical wind speeds that were regularly observed in cool luminous giants they could estimate mass-loss rates. They admitted these would be useful measures of the amount of mass loss but not quantify the mass-loss mechanism itself, which must have occurred below the dust formation zone. The~mass-loss rates they derived displayed a strong positive dependence on luminosity and negative dependence on photospheric temperature. They noted that while the values they obtained were substantial and broadly in line with those determined through other means, they fell short in some cases such as Betelgeuse, which they explained by incomplete dust condensation in those~winds.

Gehrz \& Woolf also discussed the expectations for the speed of a wind that is driven by radiation pressure on dust grains, coupled to the gas that would exhibit a reduced speed. Since Wilson \& Barrett in 1968 discovered 1612\,MHz hydroxyl (OH) maser emission from the red supergiant NML\,Cygni 
\cite{Wilson1968} it was possible to test these ideas. The~population inversion necessary for the amplified stimulated emission would be pumped by the copious emission from dust through mid-infrared transitions in the OH molecule, and~the amplification pathway would be radial through the coasting more distant parts of the wind in which water would have dissociated to supply the reservoir of OH 
\cite{Goldreich1976}. This results in a maser line profile that typically has the shape of the Minoan horns of consecration, where the peak separation corresponds to double the wind speed. With~the maser transition narrowing below the thermal width to sub-km\,s$^{-1}$ and radio spectroscopy having no issue attaining such resolution, this provides a powerful means of measuring the gas speed in the dusty wind itself. But~while OH maser emission was readily detected from around many Mira variables, assumed all to be asymptotic giant branch stars, Bowers \& Kerr discussed the paucity of red supergiant OH masers 
\cite{Bowers1978}. They did use the faster winds observed for the red supergiants to assess the plausible fraction of supergiants among the optically invisible 
OH/IR sources, which resulted in a similar number to those already known or at most 10 per cent of the OH/IR population. This may be important since those dust-enshrouded stars may embody the phase of most intense mass~loss.

It should be noted that OH maser photon flux has been used to determine mass-loss rates; however, these are dependent on the infrared flux and as such do not provide a truly independent measure from the dust-derived rates. Furthermore, besides~having the same, or~arguably worse, problem of being a trace component of the total flow of matter, the~masing process is highly non-linear and can easily break down, leading to underestimates of the mass-loss rate. But~the ability to measure the wind speed gives us a way to check whether we at least understand the way the wind is driven, and~thus reassurance of the assumptions made to derive the mass-loss rate. In~a fully coupled (allowing for drift) radiation-on-dust driven wind, the~expectation is that the wind speed depends on luminosity and the gas/dust 
mass ratio $\psi$ as $v\propto\psi^{-1/2}L^{1/4}$ since a certain fraction of the photon momentum is imparted upon the wind, $\dot{M}v\propto \tau L$ in which the optical depth $\tau\propto(\dot{M}/\psi)/(v\sqrt{L})$ (amount of dust divided by the spatial dilution factor, realising that the inner edge of the dust envelope is set by the luminosity)
\cite{vanLoon2000}. This positive correlation with luminosity and dust content was confirmed 
\cite{Marshall2004}, and~the expectation for the wind to be accelerated in the 22\,GHz water masing region from gravitationally bound ($v(r)<v_{\rm esc}(r)$) to unbound ($v(r)>v_{\rm esc}(r)$) was confirmed too 
\cite{vanLoon2001} and proven beyond doubt in the spectacular time-lapse interferometric maps obtained by Anita Richards 
\cite{Richards1998a,Richards1998b,Richards1999}.

Goldman~et~al.\ 
\cite{Goldman2017} went one step further and used the measured wind speed and luminosity to determine the gas/dust ratio. This confirmed what Gehrz \& Woolf 
\cite{Gehrz1971} suspected and van Loon~et~al.\ 
\cite{vanLoon2005} emphasised, that only the coolest red supergiants with strong pulsation achieve the full potential of dust condensation, and~that warmer red supergiants are less dusty not (just) because their mass-loss rates are lower but (also) because the gas:dust ratio in their winds is higher. This is still something that presents challenges and is not always fully appreciated in studies that present dust-based mass-loss rates. Using a radiation transfer code, grids of pre-computed models based on such codes, or~colours as proxies for the amount of attenuation (optical or near-infrared) or emission (mid-infrared) by dust, the~optical depth is determined. But~the optical depth is primarily a reflection of the shape of the spectral energy distribution, and~needs to be scaled if a mass-loss rate is to be derived from it (as stated above). This result scales with the measured or adopted values of $\psi$ and $L$, as~$\dot{M}\propto\psi^{1/2}L^{3/4}$ (see Elitzur and Ivezi\'c 
~\cite{Elitzur2001}), so if $\psi$ is actually higher than what was assumed (less dusty) then the mass-loss rate will have been underestimated; the opacity of the dust also depends on the size and composition of the grains, which can differ for sources in which dust formation is incomplete from those in which the grains are fully grown and coated. Likewise uncertainties in the luminosity will propagate, whether they originate in distance, inaccurate or lacking photometry, or~geometry (which we shall discuss later). However, if~the wind is not driven by radiation pressure on dust, coupled to the gas, which is probably the case in early-M type red supergiants without OH masers, then the scaling will instead follow $\dot{M}\propto\psi L^{1/2}v$ as one would expect from the continuity equation—assuming a constant product $\psi v$ throughout the dust envelope (though the dust could even be stagnant).

Rotational, thermal transitions of the abundant, ubiquitous carbon-monoxide (CO) molecule in the outflows from cool giants have been very useful to determine both wind speed and mass-loss rate. Their use in the case of red supergiants has been less successful, partly because the lines are not as strong and partly because they are more often contaminated by strong interstellar CO emission. Why CO is suppressed in red supergiant winds is not clear, but~
\cite{Josselin1998}  provides clear evidence for it and 
\cite{Josselin1996} suggests 
that more extensive chromospheres and/or photodissociation from the external ultraviolet radiation field may be responsible, and~they hint also at the possibility that the collisional excitation may be less effective in the cooler, more diluted dust envelopes of red supergiants. We shall return to the use of CO as a measure of mass loss~below.

Attempts to detect atomic hydrogen, which would seem an obvious thing to do, through the 21\,cm H\,{\sc i} line have been largely unsuccessful. The~intrinsic weakness of the spin-flip transition, confusion with interstellar H\,{\sc i} and the uncertain, possibly small fraction of hydrogen in atomic form (as opposed to molecular or ionised) all contribute to the challenge. G\'erard~et~al.\ 
\cite{Gerard2024} summarise the state of play, and~describe the small number of detections of almost exclusively asymptotic giant branch stars as largely reminiscent of circum--interstellar interfaces (detached shells, arcs and head-tail morphology). However, since H\,{\sc i} could potentially trace the bulk of mass loss and map the environment into which a supernova remnant would evolve, and~would complement CO in terms of dissociation-dictated envelope location, it should remain on the radar for upcoming facilities such as the new-generation Very Large Array and the Square Kilometre~Array.

We will not delve into the plethora of sub-mm lines that could be used in certain cases to trace the mass loss, though~it may well be that a robust diagnostic emerges that can be applied to samples of red supergiants. This leaves to note that radio free--free continuum emission has been detected from the ionised gas (electrons) in the chromospheres of some red supergiants such as Betelgeuse and Antares. These provide in principle a means of measuring the amount of gas surrounding the star, but~the scaling to total mass is fraught with~uncertainty.

In an attempt to offer an empirical formula for mass-loss rates that would be practical to apply to other stars, de Jager, Nieuwenhuijzen \& van der Hucht 
\cite{deJager1988} compiled estimates based on the aforementioned methods and used Chebychev polynomials in logarithmic $T_{\rm eff}$ and $L$ space. Temperatures and luminosities are also predicted by the theoretical stellar evolution models though they do depend on initial conditions (mass, metallicity, rotation) as well as the details of convection, opacities and mass loss itself. They note that it is remarkable that a single form of formula accounts for the mass-loss rates across all of the Hertzsprung--Russell diagram, with~the coolest red supergiants attaining high rates of order $10^{-6}$--$10^{-4}$\,M$_\odot$\,yr$^{-1}$. Nieuwenhuijzen \& de Jager
\cite{Nieuwenhuijzen1990} then tried to reformulate a subset of the same compilation in more physically motivated parameters as $\dot{M}=9.6\times10^{-15}L^{1.42}M^{0.16}R^{0.81}$ (all solar units and per year), though~of what they called ``fundamental stellar parameters'' only $M$ is, since $L$ and $R$ are structural and evolutionary consequences. This suggests a modest explicit dependence on mass though the positive scaling with luminosity can masquerade a stronger underlying dependency on mass. On~the other hand, mass loss could reduce the current mass but increase the size for the coolest red supergiants, increasing the mass-loss rate, until~the mantle becomes transparent and the star shrinks and becomes a yellow hypergiant, at~which point the formula suggests mass loss would~weaken.

Hagen, Stencel \& Dickinson 
\cite{Hagen1983} drew attention to 
the fact that the dustiest cool giants show weak or none of the atomic absorption so characteristic of warmer giants, and~that no maser emission is detected in the latter, thus concluding that there are fundamental differences between those winds where dust matters and those where it does not. Strong, slow pulsation, a~late-M type photosphere and 1612\,MHz OH masers together appear to be sufficient, possibly necessary, conditions for the radiation driven wind assumptions to hold true; however, this does not mean that the mass loss is driven by radiation pressure—the pulsations are more likely to be (part of) the cause. And~this brings it closer (literally) to the origin of mass loss at the base of the atmosphere where the winds of warmer red supergiants are also launched. Judge \& Stencel 
\cite{Judge1991} and also McDonald \& van Loon 
\cite{McDonald2007} showed that the smooth transition between chromospheric and pulsation, dust-aided driven mass loss compels 
the notion that processes in the star will select to grow in importance to supply the necessary impetus for mass loss whatever the type of star, suggesting we must look at more fundamental principles governing the mass loss. Whilst their studies focussed on lower-mass cool giants the insight may still also be applicable to~supergiants.

Goldberg, in 1979, proposed a scaling with surface area 
\cite{Goldberg1979} (also emphasised by Dupree 
\cite{Dupree1986}), which of course is an inverse relation with surface gravity with mass as a moderating factor as may be expected. In~this picture, photon energy loss per unit area is simply proportional to $T_{\rm eff}^4$, and~one could surmise that mass loss per unit area would also be proportional to some relevant yield parameter. This could be the inverse of the gravitational potential energy to be gained, which would bring the proportionality to $R^3/M\propto1/\rho$. Indeed, this would appear to improve the correspondence to the summary of mass-loss rates presented in Dupree's review 
\cite{Dupree1986}. It would also result in a particulate-over-radiation loss ratio of $\dot{M}/L\propto(R/M)T_{\rm eff}^{-4}$, which can be written as $\dot{M}/L\propto(L^{1/2}/M)T_{\rm eff}^{-6}$ and compared to $\dot{M}/L\propto T_{\rm eff}^{-6.3}$ derived in van Loon~et~al.\ 
\cite{vanLoon2005} for cool luminous stars with dust-driven winds. Note that the strong temperature dependence simply reflects the great variation in internal and external conditions around the boundary layer across a small range in temperature at the cool side of the Hertzsprung--Russell diagram. The~proportionality of mass loss with luminosity would then seem misleading and inspection of their ratio may bring us closer to the actual physical cause for the loss of both radiation 
and matter. For~instance, Goldman~et~al.\ 
\cite{Goldman2017} showed that red supergiants from Schr\"oder \& \mbox{Cuntz 
\cite{Schroeder2007}} exhibit a relation with surface gravity according to $\dot{M}/L\propto1/g$, which recovers Reimers' law except for their additional inverse proportionality with radius whilst Judge \& Stencel 
\cite{Judge1991}, on~the other hand, found a somewhat steeper dependence on gravity for low--intermediate-mass cool~giants.

We appear to be converging towards a realisation that mass loss from red supergiants is likely determined by a process intimately linked to gravity (or entropy), be it directly through pulsations 
\cite{Willson1986,Bowen1988} or indirectly through convection and linked to the generation of waves 
\cite{Cuntz1990,Cranmer2011}. In~a recent attempt, Kee~et~al.\ 
\cite{Kee2021} developed a formalism akin to the Parker solar wind model, in~which turbulent pressure pushes matter through the critical (turbulence-modified sonic) point after which it is lost. Whether it is obvious that the latter should happen can be debated, but~they estimate broadly similar rates to those that are inferred from observations. Their formula exhibits a very strong positive dependence of the mass-loss rate on radius but an exponential inverse dependence on mass, but 
their combination is more complicated. Since in the 
first instance the radius of a more massive red supergiant is larger, but~mass loss reduces mass and inflates the mantle, the~mass-loss rate is thus expected to display complex behaviour through the Hertzsprung--Russell diagram. The~mass-loss rate in their model depends only weakly on effective temperature but, as~expected, strongly on turbulence speed. More interestingly, it does not explicitly depend on metallicity. Fuller \& Tsuna 
\cite{Fuller2024} expanded on this work to consider how the stochastic nature of convective motion and the resulting shock waves would sustain a dynamic chromosphere from which a radiation-driven dust wind could remove matter. While their model naturally develops the dense extended atmospheres seen in the early light curves of hydrogen-rich supernovae, the~coupling with the wind is rather ad~hoc and could make it susceptible to failure in warmer red supergiants with little dust. Their mass-loss rates depend strongly on luminosity ($\approx$$L^{3.5}$) but this includes the effect of mass; in fact, their mass-loss rates are essentially set by the escape speed at the stellar photosphere, on~which they depend inversely exponentially—this is also, if~in a more convoluted manner, true for the Kee~et~al.\ model which is expressed primarily in terms of the ratio of mass and radius (potential energy, or~$gR$). In~principle this makes the mass-loss gravity linked and independent of metallicity, but~the requirement of the dust-driven wind carries an element of metallicity in it. On~the other hand, since dust-driven winds are typically found in conjunction with strong pulsation, the~latter should surely alleviate this~weakness.

Pulsation has been favoured as a launch mechanism for the coolest luminous giants because when they oscillate in the fundamental mode, like Mira variables do, the~fraction of luminous energy that is stored temporarily in the expanding mantle is very significant, and~so is the increase in atmospheric scale height 
and duration of the cycle. This could naturally create the high densities at several stellar radii distance where a dust-driven wind could form, which for overtone pulsations or convection was not obviously the case. Fundamental mode pulsation predicts a simple relation between period and gravity of the form $P\propto 1/g$ (
\cite{Wood1990}) as well as a period--luminosity relation. Motivated by this, Vassiliadis \& Wood 
\cite{Vassiliadis1993} constructed a dual regime mass-loss rate prescription in which the rate initially increases exponentially with period until it saturates because of the inability of the dust-driven wind to carry away the mass at such a high rate due to exhaustion of the radiation momentum. The~growing part of this recipe is reminiscent of the $1/g$ dependence we encountered earlier; however, the Vassiliadis \& Wood parametrisation is different (exponential, not proportional) and has incorporated the luminosity dependence implicitly within the period parameter. Furthermore, it was calibrated for asymptotic giant branch stars and admitted in need of modification (reduction) by mass. Nonetheless, Goldman~et~al.\ 
\cite{Goldman2017} successfully described the mass-loss rates of the most luminous and dustiest OH/IR stars in a form consistent with $\dot{M}/L\propto P^{3/4}\propto g^{-3/4}$ or $\sim$1/g, in~the superwind regime where Vassiliadis \& Wood would have expected an approximately constant $\dot{M}/L$.

A common misconception is that the small amplitudes of variation of red supergiants rule out pulsation as a viable mechanism to launch mass loss, in~contrast to what is widely accepted for asymptotic giant branch Mira variables. Because~amplitudes of light curves are exclusively expressed in magnitudes, they ignore the fact that these are relative quantities. A~variation in magnitude of a luminous star corresponds to a larger amplitude in radiative energy rate, $\dot{E}$ than the same variation in magnitude of a less luminous star. When van \mbox{Loon (\cite{vanLoon2002,vanLoon2008}}) examined the relation between $\dot{E}$ and wind kinetic energy rate, $\dot{K}$ they found that while red supergiants have reduced luminosity-normalised ratios $\dot{E}/L$ and $\dot{K}/L$ compared to the most extreme asymptotic giant branch stars the $\dot{K}/\dot{E}$ ratio is very similar, mildly increasing with period but generally around $\sim$$10^{-5}$. Incidentally, their $\dot{K}/L$ ratios appear similar to the luminosity-normalised chromospheric radiation-loss rates for warmer giants determined by Judge \& Stencel 
\cite{Judge1991}, suggesting they were right in emphasising radiative losses to be considered alongside material losses and providing tantalising evidence that no matter the driving mechanism, mass loss attains rates according to more fundamental, intrinsic properties and~principles.

And yet, astronomers have stubbornly expected reduced mass-loss rates at lower metallicity, probably because radiation pressure on dust, as~on metals in hot star winds, drives (some) winds. This is despite observational evidence that the only explicit metallicity ($Z$) dependence of dusty winds is their gas/dust ratio ($\psi\propto 1/Z$) and hence wind speed ($v\propto\psi^{-1/2}$) but not the total mass-loss rate ($\dot{M}\propto Z^0$), as presented and tested by van Loon 
\cite{vanLoon2000,vanLoon2006} and quantified in mass-loss rate recipes by 
\cite{vanLoon2005}, $\dot{M}(T_{\rm eff},L)$ and 
\cite{Goldman2017}, $\dot{M}(P,L)$. Winds driven by radiation pressure on dust are seen exclusively in combination with strong, slow pulsations, which in turn are only seen in sufficiently metal-rich cases that can attain the minimum dust fraction required for such wind driving. But~while Mauron \& Josselin 
\cite{Mauron2011} made an excellent comparison between carefully compiled mass-loss rates for red supergiants and the de Jager~et~al.\ 
\cite{deJager1988} prescription, vindicating the latter to broadly trace the observed trends, the~evidence they provided for a proposed scaling $\dot{M}\propto Z^{0.7}$ is extremely tenuous and could easily imply another parameter of which the variation is unaccounted for in the simplistic $\dot{M}(T_{\rm eff},L)$ parametrisation by de Jager~et~al. Comparisons between empirical estimates, and~relations based on them, are notoriously plagued by data uncertainties, some of which are systematic, and~differences in assumptions made when deriving mass-loss rates from measurements, which can be difficult to account for but also are not always valid. This becomes more problematic when they are based on disparate types of measurement or derived for disparate types of stars where the mass loss may manifest differently. In~the absence of strong evidence for metallicity dependence of red supergiant mass loss, it must be assumed there is none. Not the other way~around.

Bonanos~et~al.\ 
\cite{Bonanos2010} and more recently Antoniadis~et~al.\ 
\cite{Antoniadis2025} confirmed the earlier findings that most red supergiants in metal-poor systems ($Z<\frac{1}{2}Z_\odot$) are warmer and lack circumstellar dust compared to red supergiants in more metal-rich systems ($\sim$$Z_\odot$). Only the most extreme (cool, luminous and pulsating) metal-poor red supergiants become dusty, and~these are likely to still straddle the critical conditions for the onset of dust formation, leading to apparent or exaggerated episodic mass loss due to time-variable conditions as evidenced by 
\cite{deWit2025}. This makes it harder to compare mass-loss rates based on the effect this dust has on the spectral energy distribution, made worse by the generally unknown (and not applied) correction for the cases where dust formation has not reached maximum capacity, which lead to underestimates of the true mass-loss rates. The~latter is more likely to happen for metal-poor stars, potentially mimicking a suppression of mass loss at low metallicity. A~compounding factor is that metal-poor populations of red supergiants tend to be smaller in number, hence rare extreme phases such as the dustiest mass loss episodes are easily missed. Indeed, van Loon~et~al.\ 
\cite{vanLoon1999} showed strong evidence for a two-tier mass-loss regime for red supergiants in the LMC, with~most losing mass at rates below the nuclear mass consumption rate for core helium burning ($<$$10^{-5}$\,M$_\odot$\,yr$^{-1}$) and only a few exceeding it—but by a lot. This has since been confirmed by Verhoelst~et~al.\ 
\cite{Verhoelst2009} (in the Milky Way), Groenewegen~et~al.\ 
\cite{Groenewegen2009} (in the LMC and SMC), Bonanos~et~al.\ 
\cite{Bonanos2010} (in the SMC) and Javadi~et~al.\ 
\cite{Javadi2013,Javadi2018} (in M\,33). This would mean that either red supergiants enter a short-lived phase of intense mass loss, of~which the exact duration would be critical to determining their fate, or~only some red supergiants experience this superwind (either for intrinsic or extrinsic reasons). However, again, the~mass-loss rates at the lower end may have been underestimated due to the uncertain gas/dust ratio in those circumstellar winds, which could narrow the gap, though~is~unlikely to close it~altogether.

Studies of red supergiants in star clusters have the potential of tracing the evolution of stars with approximately the same birth mass and the same composition, and~pinpoint the dependencies on mass and composition through comparison between different suitable clusters. The~selection bias against short-lived or special and thus rare phases is even more severe than in entire galaxies, but~it could certainly help decipher the mass loss endured by most of the red supergiants most of the time. Dynamical effects and feedback from other massive stars in the cluster could affect the mass-loss process or its observational signatures, though~to some extent this would simply be part of the life of a red supergiant just like binary interaction would be in some, perhaps many or even most, cases. An~additional practical difficulty for the more populous Galactic clusters containing red supergiants is that these are generally situated in the depths of the Milky Way and are affected by absorption and emission by interstellar dust (and gas). This notwithstanding, \mbox{Beasor~et~al.\ 
\cite{Beasor2020,Beasor2023}} were able to determine mass-loss rates for red supergiants in a small sample of clusters based on dust, and~Decin~et~al.\ 
\cite{Decin2024} achieved this for one of their clusters based on CO. The~mass-loss rates they determined showed a strong dependence on luminosity, approximately as $\dot{M}\propto L^{3.5}$
as in the chromospheric model of Fuller \& Tsuna 
\cite{Fuller2024}. Emma Beasor's results were qualitatively confirmed by Humphreys~et~al.\ 
\cite{Humphreys2020}, who noted that at the low luminosity end the measured mass-loss rates fell below the de Jager~et~al.\ 
\cite{deJager1988} predictions but there is a sharp upturn in mass-loss rate above about $L>10^5$\,L$_\odot$. The~empirical values may have suffered from the same inherent uncertainty about the dust and CO fractional abundances in the outflows as described above, though, and~thereby artificially steepened the luminosity dependence by underestimating the mass-loss rates at the lower end. The~formulae presented by Beasor~et~al.\ 
\cite{Beasor2020} and Decin~et~al.\ 
\cite{Decin2024} also included an explicit dependence on mass, in~an exponential form ($\log\dot{M}\propto M$) which could be linked to the exponential dependence on escape speed in the theoretical formula derived by Fuller \& Tsuna. Note, however, that the mass in the empirical formula is that at birth whereas the actual mass will have since diminished (or increased due to binary mass transfer or merger), while the escape speed in the theoretical formula is that based on the current mass (and radius). Unfortunately, the~latter is difficult to measure directly as it requires spectroscopic gravity measurements (see 
\cite{Farrell2022}) which are compromised by atmospheric dynamics and a wavelength-dependent photospheric radius. The~advantage of pulsation-period-based 
empirical formulae such as the ones from Vassiliadis \& Wood 
\cite{Vassiliadis1993} and Goldman~et~al.\ 
\cite{Goldman2017} is that these are based on the current mass and radius, and~the application to cluster members could then provide a measurement of the mass lost up to that time (since their birth mass would be known). The~situation may be complicated by the eventuality that some of these cluster members may have undergone mass transfer with a companion or even merged, altering their masses from birth 
\cite{Wang2025}.

The mass-loss history is to some extent enshrined in the outflow, too. It can affect the overall shape of the spectral energy distribution of dusty red supergiants, but~this can be rather subtle and hard to disentangle from other effects such as departures from spherical symmetry. Modelling the relative strengths of a variety of transitions of abundant molecules arising from different parts of the outflow is a powerful, but~not often viable way to read the mass-loss history. Who else but Leen Decin~\cite{Decin2006} managed to do exactly that for the iconic Galactic red supergiant and irregularly variable OH/IR star VY\,Canis\,Majoris
, finding that this star experienced a century-long superwind at a rate of 
$\dot{M}$$\sim$$3\times10^{-4}$\,M$_\odot$\,yr$^{-1}$ a millennium ago. This confirmed the earlier findings by Smith~et~al.\ 
\cite{Smith2001}, who discovered the onset of this high mass loss 1200 yr ago on the basis of Hubble Space Telescope images. It has since calmed down by a factor of (only) four, but~came from a much more moderate $10^{-6}$\,M$_\odot$\,yr$^{-1}$, possibly proof of the suspected transition from chromospheric mass loss at sub-nuclear rates to the brief superwind phase but without explanation. \mbox{Humphreys~et~al.\ 
\cite{Humphreys1997}} used spatially resolved optical and infrared images to determine the mass-loss history of the Galactic yellow hypergiant (spectral type A--F) IRC\,10$^\circ$420. It is optically visible but still surrounded by dust and even OH masers, which is attributed to an episode of intense mass loss some half a millennium ago in which as much as a solar mass was shed at a rate of 
$\dot{M}$$\sim$$10^{-3}$\,M$_\odot$, preceded by a few dozen millennia of mass loss at a varying rate of $\sim$$10^{-4}$\,M$_\odot$\,yr$^{-1}$ during which another 10\,M$_\odot$ may have been lost. One may speculate that this star represents the final stages of the transition from a red supergiant to a yellow hypergiant, with~VY\,CMa being caught in the midst of it. Hydrogen-rich supernova light curves also indicate enhanced mass-loss episodes on centuries--millennium timescales 
\cite{Soria2025}, possibly also for lower-mass hydrogen-rich progenitors 
\cite{Folatelli2025}, though these have not been directly identified and thus their status is~uncertain.

Recently, the~extreme red supergiant WOH\,G64 has been seen to metamorphose from a red supergiant into a hotter star whilst undergoing a mass ejection (\mbox{Ohnaka~et~al.\ 
~\cite{Ohnaka2024}}). It was the coolest and most luminous red supergiant known in the LMC by a long shot, undergoing regular, long-period ($\sim$900 days) large-amplitude Mira-like brightness variations, and~boasting a huge infrared excess and the first detected extragalactic circumstellar masers 
(OH~\cite{Wood1986}; SiO~\cite{vanLoon1996}; H$_2$O~\cite{vanLoon1998}). Its mass-loss rate had been estimated to be \mbox{$\sim$10$^{-3}$\,M$_\odot$\,yr$^{-1}$ 
\cite{vanLoon1999,vanLoon2005}}. But~within a mere decade the pulsations had all but ceased and the spectrum had become that of a yellow hypergiant (Mu\~noz-Sanchez~et~al.\ 
~\cite{MunozSanchez2025}). It is not clear whether we had witnessed the ``usual'' transformation from an OH/IR star into a post-red supergiant, or~the spectacular interaction with the long-suspected B-type companion star—recently, the~Galactic yellow hypergiant HD\,144812 was found to be a post-red supergiant transferring mass onto a hot companion 
\cite{Kourniotis2025}. What is interesting though, is that all of these, VY\,CMa, IRC\,10$^\circ$420 and WOH\,G64 are thought to have descended from massive (30--40 M$_\odot$) main-sequence stars and were at least at some point in the recent past late-M type, very luminous and very dusty. These are not the kind of progenitors that have been identified for core-collapse supernovae. Instead, the~latter have been exclusively $<$20\,M$_\odot$ (see the excellent overview of this "problem" and the caveats presented by Healy~et~al.\
\cite{Healy2025}). Interestingly, Humphreys~et~al.\ 
\cite{Humphreys2020} noticed that this is also the distinction between those lower-mass red supergiants that ascend a branch, and~those massive red supergiants that merely cool 
\cite{Ekstroem2012}, which could mean that red supergiants with birth masses no more than $\sim$20\,M$_\odot$ spend so long losing mass at sub-nuclear rates that their cores collapse, whilst more massive red supergiants eject most of their hydrogen mantles and make the excursion back towards hotter photospheres 
\cite{Humphreys2023}, before~eventually their cores too, collapse—though it is not certain where exactly this mass division lies or whether indeed it is strict and does not also depend on other factors (see for instance the suggestion of an initially 16-M$_\odot$ yellow supergiant supernova progenitor 
\cite{Niu2024}). The~demarcation around $\sim$20\,M$_\odot$ affecting the advanced evolution of red supergiants coincides with a difference in carbon burning: convective, leading to smaller cores below it, and~radiative, leading to more extended cores that might implode without supernova 
\cite{Sukhbold2020}.

Briefer and less permanent changes are also observed, in~the form of eruptive and directional mass-loss episodes on timescales of year(s) in duration and typically a century in recurrence. Freshest in our memories, Betelgeuse underwent a ``Great Dimming'' in late 2019/early 2020,
perfectly placed in the evening sky—see the review by Dupree \& Montarg\`es 
\cite{Dupree2025}. This was likely due to the formation of a dust cloud above a cool convection spot—in gas originally lifted by energy deposition from a hot spot—moving in front of the photosphere, thereby attenuating it. Because~the early-M type star Betelgeuse is not efficient in producing dust in its warm chromosphere, any combination of locally enhanced density and reduced irradiation may cause such discrete dust cloud to form, which will subsequently be catapulted out. More such events will have occurred over the past century, but~missed if they did not happen or move in front of the star, or~if they happened during solar conjunction. Resolved images of the circumstellar environment of Betelgeuse show multiple structures that could have arisen from such dust bunnies. Episodic mass loss of this nature appears to be quite common 
\cite{Anugu2023,Anugu2024} and may also be happening at higher luminosities 
\cite{MunozSanchez2024}. However, we caution not to get overly excited about the significance of these outbursts, as~the sensitivity to dust formation likely amplifies the actual contrast in mass being lost through these discrete clouds vis-\'a-vis the more steady wind. Indeed, Montarg\`es~et~al.\ 
\cite{Montarges2019} discovered that in CO, the~wind of $\mu$\,Cephei (Herschel's ``Garnet Star'') is very cloudy but the smooth component dominates the mass being lost. At~the cooler, luminous end, though, these stochastic gas 
ejections may be more extreme and constitute a substantive part of the mass loss 
\cite{Humphreys2022}. Circumstellar structures attributed to once-a-century eruptions associated with individual convection cells and accompanied by dimming events have been discovered around VY\,CMa 
\cite{Humphreys2021,Humphreys2025}, so this appears to be a common phenomenon in the extended atmospheres of all red supergiants. On~longer timescales, though, when larger circumstellar volumes are sampled, measured mass-loss rates may display a more steady, tighter relation to stellar~parameters.

We finally address the elephant in the room: departure from sphericity leading to anisotropy both in the mass loss and photon loss. Already Reimers
\cite{Reimers1977} suspected that rotation may enhance mass loss, and~de Jager~et~al.\ 
\cite{deJager1988} identified rapidly rotating Be stars as exhibiting higher mass-loss rates than the generalised formula predicts, but~only if the estimate is sensitive to the equatorial density enhancement. Circumstellar geometry is not always discernible, especially in distant sources or when using optical atomic absorption lines, but~in the case of dusty envelopes it typically leads to a mismatch between attenuation of starlight and infrared emission from the dust. It can be hard to distinguish between a patchy dust shell and an equatorial enhancement, but~the strength of the ubiquitous 10$\upmu$m silicate feature arising within the dust envelope itself can break this degeneracy. If~an axi-symmetric dust envelope is seen pole-on,
then the star will be bright and so will the dust emission. It typically leads to overestimates of both the luminosity and mass-loss rate, if~not accounted for. If~it is seen edge-on,
then the dust emission will still be bright, but~the star will be dim. The~silicate feature will be relatively weak in the pole-on view but strong in the edge-on view, but~agnostic to our viewing angle in the case of a patchy shell. Masers can provide supporting evidence, though~there can be many reasons for their deviation from the canonical integrated line profiles. If~their direct mapping in interferometric observations is not feasible then duplicate kinematic components in the integrated line profiles often betray bipolar outflows, where the faster knots arise in the polar outflow and the slower ones near the~equator.

Both spectral energy distribution and masers clearly suggest bipolar, equatorially enhanced winds in the luminous Galactic red supergiant NML\,Cygni (possibly the initially most massive red supergiant known), as~well as in WOH\,G64. The~latter could be contrasted with another OH/IR star in the LMC, IRAS\,05280-6910, which could be a version of WOH\,G64 seen edge-on,
whereas WOH\,G64 was confirmed through interferometry to be seen closer to pole-on
(Ohnaka~et~al.\ 
~\cite{Ohnaka2008}). The~putative dust-enshrouded red supergiant within IRAS\,05280-6910 is all but invisible and the exceptionally strong, cool dust emission is incised by very deep silicate absorption. These geometries and the afore described post-red supergiant transitions may set the scene for what is seen in the SN\,1987A system (see  
\cite{Orlando2020}), where the exploding blue(ish) supergiant is surrounded by a ring of left-over red supergiant dust, partly sculpted by the more recent faster wind and best explained by a stellar merger. In~the case of WOH\,G64 the axial symmetry may have arisen from the influence of the companion star, and~
\cite{DeBeck2025} provides indications of a companion star shaping the wind from NML\,Cygni. Reimers 
\cite{Reimers1977} already hinted at binarity enhancing mass loss, even before any merger occurred, though~Antoniadis~et~al.\ 
\cite{Antoniadis2024} show that this may only be a minor effect for most of the red supergiants. While it was not possible here to fully explore the effects of binary interaction on the evolution of massive stars (see, for~instance 
~\cite{Eldridge2017}) and their mass loss, there does appear to be evidence that binary interaction may be more important among the more luminous, more massive red supergiants and that this might contribute to, or~even be responsible for, the~superwind phase these stars~experience.

\begin{figure}[H]
\includegraphics[width=0.99\textwidth]{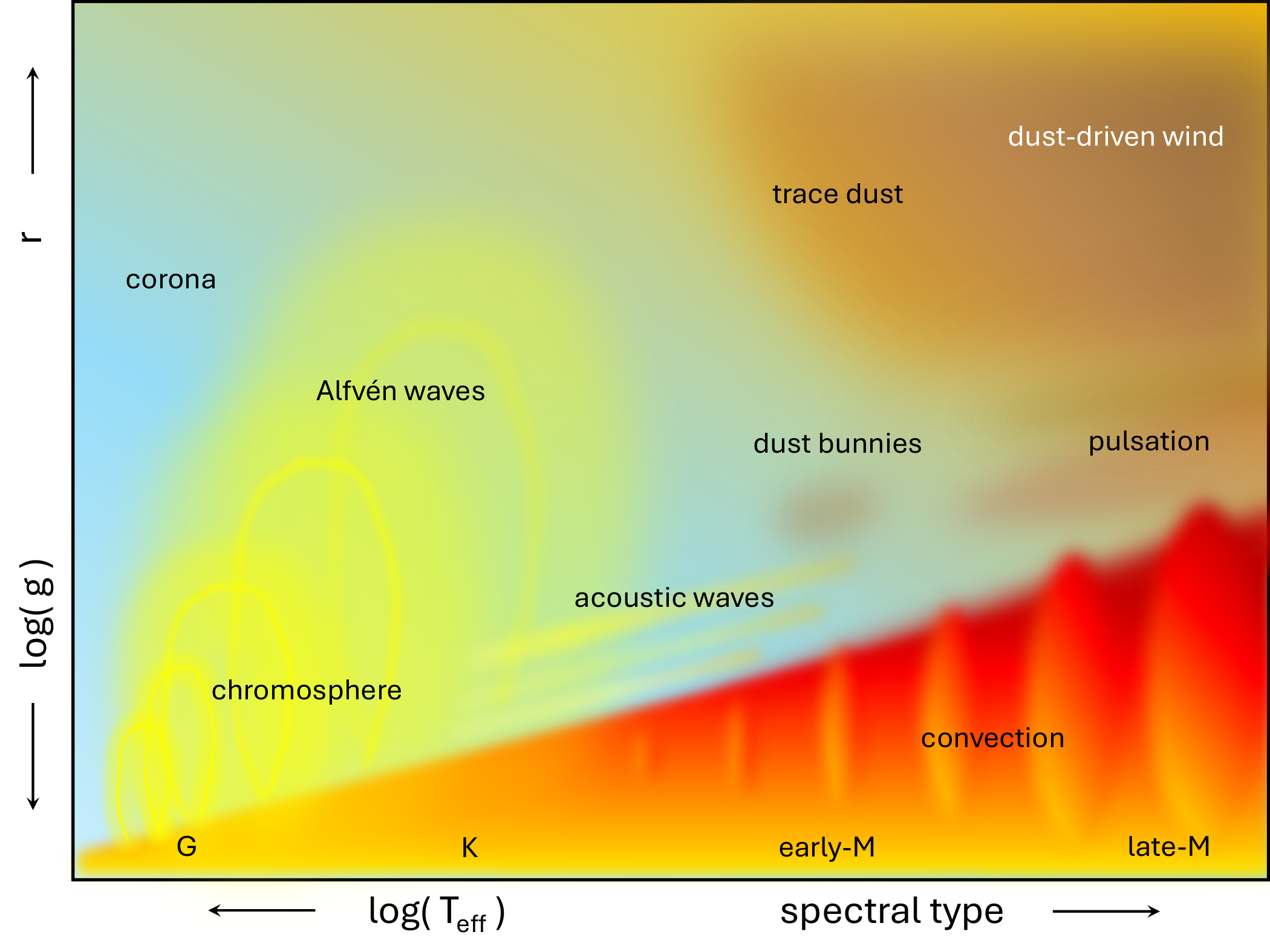}
\caption{Summary of varying conditions of the boundary layer of cool supergiants, from~the base through to the wind, expressed across stellar photospheric effective temperature and local gravity (a function of current mass and radial distance). Metal-poor stars will struggle or fail to occupy the top-right corner of this diagram but they will still—or instead—lose mass through the other means. Note that around K/early-M spectral type chromospheres are important but the sketch has separated them from the dusty regime for clarity. Betelgeuse is a prime example of where they co-exist. This diagram can easily be extended towards the left with line radiation-driven winds of OBA-type~stars.\label{fig1}}
\end{figure}

We finish by capturing some of the spatial and evolutionary aspects of the various regimes of mass loss, in~Figure~\ref{fig1}, as~a novel way in which to visualise and contemplate what happens around the boundary layers of cool supergiants, realising that metal-poor stars may not instantiate the phenomena at the lowest temperatures and~gravities.

\unskip

\section{Take-Away Points and the Road~Ahead}

On the basis of physical arguments and empirical facts, one may conclude that mass loss, like radiation loss (luminosity), is the product of gravity and not inherently dependent on metallicity. It is both produced, as~a natural consequence of a move towards equilibrium, and~constrained in a self-regulatory manner. At~the outer boundary set by the decoupling of radiation from matter, according to the local values for $g$ or $v_{\rm esc}$ almost all of the initially generated luminosity escapes but a small fraction is taken away by the mechanical energy associated with the mass loss (far exceeded by the rest-mass energy loss). Mass is lost more easily when the outer boundary has low (negative) gravitational potential energy and is therefore limited by where the optical depth is unity. Hence one expects $\dot{M}\propto L$ and any further dependence to be due to a more fundamentally gravity-related parameter in practice expressed either in terms of $g$ or $M$ and $R$ or by proxy in the form of additional terms of $L$ and/or $T_{\rm eff}$. The~fact that various mass-loss rate recipes agree on a positive dependence on luminosity but not in the exact value of the exponent (for instance, refs.
\cite{Vanbeveren1998,Salasnich1999} find $\dot{M}\propto L^\alpha$ with $\alpha=0.8$ and $2.1$, respectively) indicates the difficulty in accounting for these parameters. Yang~et~al.\ 
\cite{Yang2023,Wen2024} found a more nuanced luminosity dependence, but~it is hard to understand the physical basis for fitting a third-order 
polynomial in $\log(L)$ to the mass-loss rates. Likewise, it is not obvious what to make of temperature dependencies as steep as $T_{\rm eff}^{-33.7}$ or even about half that as presented by Antoniadis~et~al.\
\cite{Antoniadis2024,Antoniadis2025}.

Individually measured mass-loss rates are always uncertain due to the limited amount and accuracy of data, variability of the star's outer boundary ($10^{-1}$--$10^1$ yr) as well as stochastic, eruptive and episodic variations in the mass loss ($10^0$--$10^4$ yr), and~assumptions about things like plasma conditions, elemental and chemical abundances and dust properties and condensation fractions as well as the adoption of a simplified physical model. These generally add scatter to trends in $\dot{M}$ with stellar parameters, but~can also cause systematic shifts or introduce artificial trends and explain—but not always easily resolve—discrepancies between different studies. If~stars are amenable to multiple methods of deriving mass-loss rates this can reduce scatter or mitigate offsets, but~errors can be correlated especially when the methods are inter-related or have comparable assumptions in common. The~`kink' (upturn) in the luminosity dependence of the mass-loss rate as first reported by 
\cite{Humphreys2020} and subsequently by~\cite{Yang2023,Antoniadis2024,Antoniadis2025} may be due, in~part or in whole, to~a decrease in gas/dust ratio as more luminous red supergiants in those samples are cooler and consequently dustier; this is consistent with the kink occurring at higher luminosity at lower metallicity, since metal-poor red supergiants are warmer and therefore less dusty. Correcting for this trend in dust/gas ratios would make the $\dot{M}(L)$ relation less steep overall and potentially lose its kink. This is an example where empirically as well as conceptually our view of mass loss is distorted by ill-constrained 
detail. In~general, for~any method being applied to a sample of stars, the highest measured mass-loss rates are more likely to be overestimated, and~the lowest are more likely to be underestimated, sometimes severely so, leading to steeper empirical relations than the true, physical ones. While the 1970s Reimers law does not necessarily represent the ground truth, it is relatively shallow, simple and easily relatable to both luminosity and gravity and thus continues to provide a valuable benchmark for other formulae. Ultimately, though, mass loss manifests itself in relation to its physical state and evolution, which are a function of its birth mass, metallicity, rotation, binarity and time, and~efforts should focus on reducing observed behaviour to these dependencies and their root~causes.

Both the observed mass loss and the inference from evolutionary end phases strongly indicate that (initially) low-mass red supergiants do not lose mass at sufficiently high rates to deplete their mantles before undergoing core collapse, whilst the (initially) most massive red supergiants invariably do and become warmer hypergiants before core collapse occurs (cf.\ 
\cite{Zapartas2025}). Lower-mass red supergiants are not expected to become very dusty, making them easier to identify in pre-explosion images and their bolometric luminosity and hence birth mass to be determined with confidence. The~majority of core-collapse supernovae are hydrogen-rich
\cite{Ma2025}, which supports the idea that the lower end of the red supergiant mass range ($\sim$8--20\,M$_\odot$ at birth), which comprises the majority of red supergiants, are their principal forebears (see for an independent inference 
~\cite{Fang2025}) whilst the fewer more massive red supergiants ($\sim$20--40\,M$_\odot$ at birth) ultimately explode as hydrogen-depleted supernovae. Supernova ejecta masses may in fact fall short of increasing proportionally to progenitor mass. The~circumburst medium may reflect a dense and structured atmosphere, even—perhaps especially so—for low-mass progenitors, and~more extended for more massive progenitors, and~we caution not to interpret this in terms of actual mass loss at $\gg$$10^{-2}$\,M$_\odot$\,yr$^{-1}$ and winds with speeds $\gg$ 10\,km\,s$^{-1}$ as is commonly done (such as \mbox{in 
\cite{Xiang2024,Zhang2024,Bostroem2024,Zhao2024,Iwata2025,Andrews2025,Hu2025,Dickinson2025,Rehemtulla2025,Hinds2025,Jiang2025,Goto2025}} where there is typically evidence for a dense inner atmosphere and less dense outer wind)—something that may be starting to be acknowledged 
\cite{Matsuoka2025b}. Unless~greatly enhanced mass loss can be triggered by the onset of nuclear carbon burning in the final stages of the star's life, witnessing such eruption for other reasons (including binary interaction) just prior to core collapse should be highly uncommon. These predictions can all be tested through supernova progenitor demographics, ejecta mass estimates and light-curve analysis 
\cite{Dessart2025,Wasserman2025,Matsuoka2025a}. Interestingly, radio non-detections of supernovae constrain prior mass-loss rates to be generally low 
\cite{Sfaradi2025}, possibly commensurate with what is seen for typical red supergiant progenitors of modest birth mass, though~the occasional inference of $\dot{M}<10^{-4}$\,M$_\odot$\,yr$^{-1}$ in the final $\sim$10$^3$\,yr before explosion is plausible 
\cite{JacobsonGalan2025}.

Binary interaction is likely to enhance mass loss, not suppress it, and~may affect the more massive systems more, not least since these red supergiants become larger and more tenuous; however, this may in reality be more mixed 
\cite{Solar2024,Nagao2025}. This will have a stochastic effect on a population scale and could lead to dichotomy within supernova types, not just between hydrogen-rich and hydrogen-poor. This, too, could be tested but would require a finer analysis and theoretical understanding of supernova characteristics (including 
\cite{Ercolino2024,Dessart2024,Rose2024,BaerWay2025,Reynolds2025}) and benefit from a joint analysis of post-red supergiant survivors. In~this respect it is interesting to see the estimated birth masses of warm progenitors of type IIb supernovae ($\sim$16--18\,M$_\odot$, see 
\cite{Niu2025}) hugging just below the upper mass limit ($\sim$20\,M$_\odot$) of type II-P red supergiant progenitors; these include cases like SN\,1993\,J, believed to have experienced binary interaction. Stripped-envelope supernova progenitors, purported to arise from binary interaction, seem to have systematically much higher pre-explosion mass-loss rates (or rather, circumstellar densities) 
\cite{Matsuoka2025b}. X-ray observations are consistent with modest mass-loss rates for the lower-mass red supergiant progenitors of type II-P supernovae and higher mass-loss rates for those that may have experienced a superwind (type II-L) or binary interaction (type IIb) or eruptions of some kind (type IIn) 
\cite{Dwarkadas2025}. We also draw attention to the fact that the final mass at the time of explosion can be determined from photometry quite well for warm hypergiants but not for cool red supergiants 
\cite{Farrell2020}, making it hard to determine how much mass type II-P progenitors may have~lost.

Indeed, a~holistic approach is both needed and possible, given the wealth and variety of constraining data (if still incomplete, uncertain and open to interpretation). It should combine measurements of mass loss and of stellar parameters of red supergiants, as~well as measurements of their progenitors (main-sequence and blue supergiants) including multiplicity demographics and of their progeny (yellow hypergiants, Wolf--Rayet stars, supernovae, compact remnants and so on), but~at the same time be constrained by physics as we understand it, in~a Bayesian manner. This is the only way to circumvent (and elucidate) the flaws and biases, both in the data and their analysis as well as the theoretical models for stellar evolution, atmospheres and the details of mass loss and stellar winds (and the consequences of core collapse and any eruptive behaviour during carbon burning prior to an ensuing supernova). It must account for all uncertainties. It should further be constrained by similar data and models for lower-mass stars that climb the red and asymptotic giant branches, of~which some of the physics is similar but the statistics are much better and not all trends are traced in exactly the same manner and thus provide a nuanced perspective, as~well as by those (more) massive stars that do not become red supergiants. It must be self-consistent, between~stellar mass formation and stellar mass loss—a zero sum game when accounting for supernova ejecta and compact \mbox{remnants 
\cite{Javadi2013,Javadi2018}} and imprinted upon the red supergiant and yellow hypergiant luminosity functions 
\cite{DornWallenstein2023}.

Such multi-dimensional, multi-faceted probabilistic analysis is now becoming possible with the advent of sophisticated machine learning approaches. Physicists in modern times have always aimed at dimensionality reduction and regression in order to deduct the smallest set of simplest laws required to describe reality, expressed in terms of the most fundamental accessible parameters and their combinatory and causal relationships. But~due to the phenomenological complexity combined with the unsurmountable limitations of observational astronomy,
the problem of stellar mass loss has eluded human capacity to solve it. Artificial intelligence is far superior in performing such massive inductive analysis, and~can 
uncover hidden variables such as entropy that hitherto have escaped our attention or been impossible to pin down precisely. Artificial intelligence will not immediately make redundant human effort in the acquisition of new data or the building of theoretical models, though~it will increasingly be able to direct observational strategies and modify simulations. It will still require a good deal of philosophy in order to interpret and understand what any artificial intelligence would come up with, but~machine learning will help us cut out the tedious and complex tasks, reduce bias and prejudice and allow us to concentrate on the application of human thought. It will require a substantial programme of work involving a large team of diverse minds, but~one only has to think of the first decade of human space flight, the~hunt for the Higgs boson or the deciphering of the human genome to find precedents of this nature and magnitude, though~probably at a fraction of the costs of those monumental achievements. The~future can be bright, if~we want it to~be.

\vspace{6pt} 

\funding{This research received no external funding (for quite some time).}

\acknowledgments{We are grateful to the Editor of this Special Issue, Professor Roberta Humphreys for their trust and patience. We dedicate this review in memory of Professor Paola Marigo, for~inspiring us by their faith in our ability to understand and apply physics as well as by their humanity. We thank Roberta Humphreys, Andrea Dupree and two anonymous reviewers for their positive reception of the manuscript and for their helpful comments that were all implemented in~full.}

\conflictsofinterest{The author declares no conflicts of~interest.} 

\begin{adjustwidth}{-\extralength}{0cm}

\reftitle{References}

\PublishersNote{}
\end{adjustwidth}
\end{document}